# Inside the Working Mechanism of Meta-generalized Gradient Density Functional Approximations: The Example of Quantum Spin-Hall Insulator 1T'-WTe$_2$

Li Yin[*], Hong Tang, and Adrienn Ruzsinszky

*Department of Physics and Engineering Physics, Tulane University, New Orleans, Louisiana 70118, USA*

[*]Author to whom all correspondence should be addressed.

E-mail: lyin2@tulane.edu

# ABSTRACT

Quantum spin Hall (QSH) insulators have attracted intensive experimental and theoretical studies due to their beneficial applications in spintronic devices. Density functional theory (DFT) meets challenges when describing the electronic structure of QSH materials. Only the Heyd-Scuseria-Ernzerhof (HSE06) with spin-orbit coupling (SOC) is effective in revealing the band opening in the typical QSH 1T'-WTe$_2$, but with increased computational demands. Here, using DFT, Wannier function simulations, the screened hybrid HSE06 functional, and first-principles-based many body perturbation theory $GW$, we investigate the sensitive electronic structure in monolayer 1T'-WTe$_2$, with advanced meta-generalized gradient (meta-GGA) density functional approximations. The success of the recent strongly constrained and appropriately normed SCAN and its revision r$^2$SCAN meta-GGAs left their predecessor "meta-GGA made very simple" (MVS) ignored by the scientific community. Largely unnoticed were the increased band gaps of MVS compared to any semilocal approximation including the SCAN meta-GGA. We find that the non-empirical MVS approximation yields a positive fundamental band gap, without any help from exact exchange, Hubbard $U$, or SOC correction. We explain the success of the meta-GGA MVS for the band gap in 1T'-WTe$_2$ by presenting two working mechanisms in meta-GGA approximations. Besides, we point out the difficulty of using $G_0W_0$ for this specific 1T'-WTe$_2$. Although the single shot $GW$ correction with an MVS reference yields a smaller band gap than $GW$ with PBE reference, the $G_0W_0$@MVS is still not suitable for simulating 1T'-WTe$_2$, due to its negative band gap. These DFT and beyond DFT results highlight the importance of meta-GGAs, especially the MVS approximation, and of novel construction schemes with enhanced kinetic energy density dependence. The MVS approximation re-appears as an appealing alternative for accurately describing 1T'-WTe$_2$, paving an efficient way for exploring other two-dimensional QSH materials in high-throughput calculations.

# ◼ INTRODUCTION

Quantum spin Hall (QSH) effect in the Hall effect family has been investigated over decades for its potential applications in energy-efficient quantum electronic and spintronic devices [1–5]. After the first theoretical predictions [1,2,6], experimental evidences of the QSH effect were observed soon in the typical HgTe/CdTe[3] and InAs/GaSb/AlSb[7] quantum wells. Different kinds of QSH-based devices are further proposed, where the large band gap, large number of conducting channels, and efficient on-off switching method are three critical elements for practical applications [8]. However, the fundamental band gap in QSH insulators is generally not large enough. Searching large-gap (~0.1 eV) QSH insulators becomes naturally a next step. Several two-dimensional materials, such as the thin films[9] and transition metal dichalcogenides $MX_2$ (M=W, Mo, X=Te, Se, S) in the 1T' phase [8], have been proposed as large-gap QSH insulators. The $MX_2$-based QSH devices are predicted to exhibit enhanced conductance and rapid on-off switching via topological phase transition[8]. Furthermore, the monolayer 1T'-$WTe_2$ has been demonstrated to be QSH via both first-principle calculations and experimental observations[10,11].

However, so far, the fundamental band gap in monolayer 1T'-$WTe_2$ can only be correctly revealed by the Heyd-Scuseria-Ernzerhof (HSE06) hybrid functional with spin-orbit coupling (SOC)[10,11]. Within HSE06, the semi-local PBE density functional approximation is combined with the screened Hartree-Fock exact exchange[12,13]. Neither the common density functional approximations nor *GW* correction can open the band gap in 1T'-$WTe_2$[8]. A similar phenomenon also exists in 1T'-$MoTe_2$[8]. However, the hybrid functional brings huge computational demands. Also, since the band inversion in 1T'-$WTe_2$ already happens without SOC, proper density functional approximations should be possible to describe the peculiar band structures and open the fundamental band gap in 1T'-$WTe_2$. Advanced meta-generalized gradient approximations (meta-GGAs) are considerable alternatives.

In the Perdew-Schmidt hierarchy[14], meta-GGAs are on the third rung owing to the additional ingredients of non-interacting kinetic energy density or Laplacian dependence of the local electron density[15–17], which make meta-GGAs more flexible to satisfy different constraints simultaneously as compared with common GGAs. A dimensionless ingredient defined by the kinetic energy density[15,16,18] is introduced in the best meta-GGAs to recognize slowly varying densities, one-electron regions and nonbonded interactions. To date, various kinds of meta-GGAs have been utilized to investigate different properties, such as reliable phonon dispersions in systems with distinct chemical properties[19], magnetic ground state[20], magnetic form factor[20] and metal-insulator transition[16] in high-temperature superconductors. These successes of meta-GGA inspired us to investigate the typical QSH 1T'-WTe$_2$ with advanced meta-GGAs, in order to accurately and efficiently simulate its electronic structure, especially the fundamental band gap.

In this work, using density functional theory (DFT), Wannier function simulations, screened hybrid functional HSE06, and many body perturbation theory *GW*, we analyze the electronic structure of the QSH insulator monolayer 1T'-WTe$_2$, along with the common PBE-GGA and various advanced meta-GGAs. We find that certain non-empirical meta-GGAs can describe the positive band gap in 1T'-WTe$_2$ without exact exchange, *GW*, Hubbard *U* or SOC correction. The MVS meta-GGA-calculated band gap can reach 81 meV together with SOC correction, which is close to the ARPES-measured band gap of 55±20 meV and HSE06-SOC-based gap of 69 meV. The meta-GGAs utilized in this work represent various construction designs. Along with the results, we explain the working mechanism of the meta-GGAs and their impact on the band gap and electronic structure of our workhorse material 1T'-WTe$_2$. Our results indicate the distinguished potential of meta-GGAs for the electronic structure of monolayer 1T'-WTe$_2$ and other two-dimensional materials with challenging or ultra-sensitive electronic structures.

## ■ METHODS

First-principles calculations were performed using DFT in the Vienna Ab-initio Simulation Package (VASP)[21,22], with the projector augmented wave pseudo-potentials[23,24], the Perdew-Burke-Ernzerhof (PBE) functional of the GGA rung[25], and meta-GGAs[15,16,18]. The energy cutoff of 520 eV is used for the plane waves. The valence electrons of W and Te atoms are in the $6s^15d^5$ and $5s^25p^4$ (The electronic configurations in the atoms can be used for PAW construction, but not otherwise in calculations for solids) states respectively for both the GGA and meta-GGA calculations. Here, we considered the recent meta-GGAs made simple 2 (MS2)[26,27], made very simple (MVS)[15], the strongly constrained and appropriately normed (SCAN)[18,20], the regularized-restored SCAN (r$^2$SCAN)[19,28], TASK, and modified TASK (mTASK) approximations[29,30]. The dimensionless kinetic-energy-density-based ingredient is utilized in each approximation[15,16]. The Brillouin zone is sampled with the Γ-centered 14×6×1 *k* point mesh. The lattice constants and atomic coordinates are fully relaxed in the PBE and SCAN approximations respectively, where SOC is also included. The relaxed lattice constants of monolayer 1T'-WTe$_2$ are $a$=3.486 Å, $b$= 6.326 Å in the PBE+SOC calculation and $a$=3.483 Å, $b$= 6.310 Å in the SCAN+SOC calculation, which are close to the reported values of $a$=3.503 Å and $b$= 6.311 Å[10]. In this work, we apply the optimized structure by SCAN+SOC calculation for further analysis, as shown in Figure 1a. A vacuum region of 35 Å is employed in the $c$ direction for simulating the two-dimensional nature of 1T'-WTe$_2$. The convergence criteria are $10^{-6}$ eV for the energy and 0.001 eV/Å for the atomic forces, respectively.

Single-shot $GW$ calculations ($G_0W_0$) -of many-body perturbation theory- are applied together with PBE and MVS reference approximations, respectively. The valence electron configurations of $5s^25p^65d^6$ in W atom and $5s^25p^4$ in Se atom are used for the $GW$ calculations. To reduce the computational demand, the band structure with $GW$ correction is calculated using the projected Wannier function method[31]. The W $d$ and Te $p$ orbitals are projected, without the maximal localization in Wannier90[32]. The crystal structure is visualized using the VESTA 3.4.7 code[33]. The band structures are plotted using the Gnuplot 4.6 code[34].

## ■ RESULTS AND DISCUSSION

**Performance of Meta-GGAs for the Fundamental Band Gap.** The fully relaxed monolayer 1T'-WTe$_2$ is displayed in Figure 1a, where the W and related Te atoms are distorted with the band inversion. The detailed band structures of 1T'-WTe$_2$ are provided in Figure 1b from the PBE functional, where the band inversion caused by 1T' structure distortion occurs near the Γ point (please see the $k$-path of S-Γ-X). However, as marked in Figure 1b, the supposed conduction band minimum (CBM) is found lower than the valence band maximum (VBM), generating a negative indirect band gap (defined by CBM-VBM) of -146 meV without SOC and -120 meV with SOC, which is close to the gap value of -110 meV reported with previous GGA calculations[8]. The SOC correction is able to separate the crossing between valence and conduction bands, but can't fix their energy overlapping, i.e., the VBM is still higher than CBM. It's clear that the generalized GGA approximation cannot capture the small band gap of 1T'-WTe$_2$. Next, we harness the more advanced meta-GGA-rung approximations for quantitative evaluation, and compare their performance with the hybrid HSE06 functional.

We calculated the fundamental band gaps of the 1T'-WTe$_2$ with the screened hybrid HSE06, PBE, and different meta-GGAs respectively, as summarized in Figure 1c. The corresponding energy positions of VBM and CBM are illustrated in Figure 1d,e. As compared to the common PBE exchange correlation functional, the hybrid HSE06 does mitigate the overlapping of VBM and CBM, but still leads to a the negative band gap without adding SOC.

In terms of meta-GGAs, there are mainly two kinds of approximations[15,16,18,26,28,29]. Generally, the exchange-correlation energy in meta-GGAs is defined as

$$E_{xc}^{sl}[n_\uparrow, n_\downarrow] = \int d^3 r\, n(r)\varepsilon_{xc}(n_\uparrow, n_\downarrow, \nabla n_\uparrow, \nabla n_\downarrow, \tau_\uparrow, \tau_\downarrow). \qquad (1)$$

In equation (1), $n_\uparrow$ and $n_\downarrow$ are the electron densities in spin-up and spin-down channels, $\nabla n_\uparrow$ and $\nabla n_\downarrow$ are the local gradients of the spin densities, $\tau_{\uparrow,\downarrow} = \frac{1}{2}\sum_\mu |\nabla \psi_{\mu\uparrow,\downarrow}|^2$ is the kinetic energy density constructed from the occupied Kohn-Sham orbitals $\psi_\mu$, and $\varepsilon_{xc}$ is the approximate exchange-correlation energy per electron[15,16,18]. Some recent meta-GGA approximations use the ingredient that characterizes the dimensionless deviation from a single orbital shape,

$$\alpha = \frac{\tau - \tau^W}{\tau^{unif}} \qquad (2)$$

where $\tau^W = \frac{|\nabla n|^2}{8n}$ is the von Weizsäcker kinetic energy density and $\tau^{unif} = \frac{3(3\pi^2)^{2/3} n^{5/3}}{10}$ is the kinetic energy density limit of uniform electron gas. The exchange energy can be re-defined with the inhomogeneity ingredients as

$$E_x^{sl}[n] = \int d^3 r\, n(r)\varepsilon_x^{unif}(n) F_x(s, \alpha), \qquad (3)$$

where $\varepsilon_x^{unif}$ is the exchange energy per electron for the uniform electron gas, $s = \frac{|\nabla n|}{2(3\pi^2)^{1/3} n^{4/3}}$ is the reduced density gradient, and $F_x$ is the exchange enhancement factor[15]. For the first groups of meta-GGA, including MS2, MVS, SCAN, and r$^2$SCAN, the enhancement factor $F_x$ is built to give accurate ground state properties. The second category, comprising TASK and mTASK, is constructed to yield more spatial nonlocality in the exchange component.

In monolayer 1T'-WTe$_2$, as shown in Figure 1d, without SOC correction, most of the meta-GGAs except the MVS approximation yield negative fundamental band gaps. Especially, among the first category of meta-GGAs, the VBM gradually drops down while the CBM rises, from MS2, SCAN, r$^2$SCAN, to MVS, while the band gap becomes positive with MVS approximation. Such a trend also applies to the case with SOC included, as shown in Figure 1e. Moreover, with means of SOC correction, the HSE06-calculated band gap also becomes positive and reaches 69 meV. The MVS-based band gap with SOC increases to 81 meV. These values are close to the experimental band gap of 55±20 meV according to the overall band structure and second derivative spectra of 1T'-WTe$_2$ measured by angle-resolved photoemission (ARPES)[11]. The MVS-calculated gap matches the accuracy of HSE06. The detailed band structures calculated by MVS and HSE06 - without and with SOC- can be found in Figure S1 of the supplement. One notable item is that the value of MVS-calculated band gap without SOC is illustrated as 10 meV in Figure 1c and S1a. Such a small gap can be sensitive to the density of k points from Γ to X. Thus, we further simulate that band structure using projected Wannier functions, where a better interpolation between k points is available. Figure S2 of the supplement shows that the band structures calculated by DFT and the Wannier function method are consistent with each other, and only a minor difference occurs in the VBM and CBM. According to the Wannier simulation results, although the MVS-based band gap might be less than 10 meV, the VBM is lower than CBM, denoting a positive band gap. Besides, the VBM around the Γ point gradually becomes "flat" as the approximation varies from PBE, HSE06 to MVS approximation, as shown in Figure 1b and S1. This "flat" feature in VBM also was revealed in previous ARPES results[11]. Overall, in view of the high computational cost of hybrid HSE06, the meta-GGA MVS approximation could be a suitable density functional approximation for describing 1T'-WTe$_2$.

In general, the meta-GGA performance exceeds that of GGA because of the additional introduced kinetic energy density $\tau_{\uparrow,\downarrow}$ to distinguish the iso-orbital and nearly-uniform electron gas regions, which enables meta-GGAs to satisfy more exact constraints without relying on fitted parameters. We begin our analysis of the working mechanism of meta-GGAs by comparing the dimensionless kinetic energy density variable $\alpha$. According to equations (2) and (3), $\alpha$ is essential to the construction of exchange enhancement factor $F_x$. As an essential ingredient of meta-GGAs, $\alpha$ controls spatial nonlocality. Thus, in Figure 2, we depict the distribution of $\alpha$ for monolayer 1T'-WTe$_2$ with different functionals. The (-1 1 -8) plane displayed in Figure 2a is specifically chosen since both W and Te atoms lie in this plane. Figure 2b shows that, the $\alpha$ distributions in different functionals are similar, except the circled region between Te atoms. As the approximation varies from MS2, SCAN, r$^2$SCAN, to MVS, $\alpha$ gradually increases in the circled region, reflecting a decreased electron localization[35,36] with more d-electron character between neighboring Te atoms. Meanwhile, the band gap increases from negative to positive in the same order, as shown in Figure 1c. Such a trend implies a charge transfer between *p* and *d* orbital electrons in 1T'-WTe$_2$. However, due to the strong orbital overlap, $\alpha$ is larger for *d* than for *p* regions. Thus, it's hard to capture the detailed change of *d* electrons via analyzing the $\alpha$ value only. For quantifying the change of *p* and *d* electrons in different approximations, we analyze the occupations and charge densities in the following section.

**Analysis of the Occupation and Density.** First, we calculate the atomic occupations. To do this, we normalize our spin-up Bloch states $\psi_{n\mathbf{k}}(\mathbf{r})$ from band *n* with wavevector $\mathbf{k}$ in a supercell. We put our origin on the nucleus of an atom *a* in the supercell. Then we can find what fraction of the *n*-th band is occupied from

$$O_n^a = \sum_{\mathbf{k}} f_{n\mathbf{k}} \sum_{l=0}^{\infty} \sum_{m=-l}^{l} |\langle Y_{lm} | \psi_{n\mathbf{k}} \rangle|^2 = \sum_{\mathbf{k}} f_{n\mathbf{k}}. \tag{4}$$

The first equality is a definition, and second equality follows from the completeness of the spherical harmonics for the description of any possible angular dependence about any origin: $\sum_{l=0}^{\infty} \sum_{m=-l}^{l} |Y_{lm}\rangle\langle Y_{lm}| = 1$. Note that completeness requires summing up to $l = \infty$. The final expression is the number of spin-up electrons from the *n*-th band in the supercell, which is 1 for a filled band and less than 1 for a partly-filled band. Note that it is independent of *a*. When we restrict the maximum *l* to 2, we achieve completeness only for the atom *a*. Presumably we can almost restore completeness by summing over all 6 atoms in the $W_2Te_4$ supercell, and then we get approximately the number of spin-up electrons on atom *a* from band *n*. In Figure 3, we show the average over the two W atoms, and the average over the four Te atoms, for the top valence and bottom conduction bands. The occupation is confirmed by the fact that $2O_{VB}^W + 4O_{VB}^{Te} \approx 1$.

As shown in Figure 3a, there are three features in the change of atomic occupation as follows,

Feature i) As compared with PBE, the hybrid HSE06 and the meta-GGA MVS approximation yield smaller occupation of W atoms and larger occupation of Te atoms in the band with VBM. Except for HSE06 and MVS, minor occupation of W atoms diffuses into the band with CBM. Meanwhile, as shown in Figure 1c, the HSE06- and MVS-calculated band gaps are larger than the gap from PBE.

Feature ii) Among the first kind of meta-GGAs, as the approximation varies from MS2, SCAN, r$^2$SCAN to MVS, the W occupation gradually decreases but the Te occupation increases. At the same time, the band gap shown in Figure 1c increases in the same order of the above listed approximations. Combining features i) and ii), one can state that the band gap can be improved via decreasing W and increasing Te occupations.

Feature iii) However, Figure 1c shows that, the second kind of meta-GGAs, comprising TASK and mTASK, also produce larger band gaps than the PBE and SCAN functionals. However, in Figure 3a, the W occupation with mTASK or mTASK approximation is larger than that in PBE, while the Te occupation is slightly lower than that in PBE. Such characteristic is contrary to physics observed with HSE06 and the first kind of meta-GGAs, suggesting a different mechanism of these meta-GGAs for band gap prediction.

Feature iv) For all the functionals, an additional SOC correction can enlarge the band gap, along with the increased W occupation and decreased Te occupation, as shown in Figure 1c and 3. The SOC effect and the feature iii) suggest that the band gap in 1T'-WTe$_2$ can also be enlarged via increasing W and decreasing Te occupations, but using a different kinetic energy density dependence in $F_x$. Moreover, feature ii) also includes SOC, i.e., considering the SOC correction, the W occupation still decreases but the Te occupation increases in the approximation order from MS2, SCAN, r$^2$SCAN to MVS, as shown in Figure 3b.

So far, we have found two mechanisms to improve the band gap in 1T'-WTe$_2$. One is decreasing W and increasing Te occupations, as the HSE06 and first kind of meta-GGAs do. On the contrary, another mechanism is increasing W but decreasing Te occupations, as the second kind of meta-GGAs and SOC correction do. We note that another mechanism of meta-GGA functionals for band gaps based on the description of the highest occupied and lowest unoccupied orbitals in the space of meta-GGA ingredients was introduced recently[37]. For further clarifying the two mechanisms, we plot the spatial distribution of partial charge density in 1T'-WTe$_2$ (see Figure S3). Considering that only minor atomic occupation diffuses into the band with CBM, we focus on the partial charge density in the band with CBM.

As shown in Figure S3 of the supplement, we plot the partial charge density from the band with the VBM on the two-dimensional (001) plane of W atomic layer. The HSE06-calculated charge densities between W atoms are smaller than that calculated by the PBE functional. As the approximations vary from MS2, r$^2$SCAN, SCAN, to MVS, the charge density between W atoms gradually decreases. The TASK- and mTASK-calculated charge densities between W atoms are larger than that in PBE. All these features are consistent with the atomic occupations revealed in Figure 3a. Moreover, according to Figure S3, the functional-caused charge density differences between W atoms are most obvious at the density value around 0.0035 e/Bohr$^3$, corresponding to the cyan region in the color scale. Therefore, we plot the partial charge density at the specific isosurface of 0.0035 e/Bohr$^3$ in Figure 4, which explicitly exhibits the spatial distribution.

As shown in Figure 4, for the first mechanism revealed by HSE06 and the first kind of meta-

GGAs, the decreased W and increased Te occupation correspond to the decreased *d-d* interaction between W atoms (see the decreased charge densities in the dashed circle) and increased *d-p* orbital hybridization between W and Te atoms in the band with VBM (see the increased charge densities in the dashed rectangle). As the approximation varies from MS2, SCAN, r$^2$SCAN to MVS, decreased *d-d* interaction and increased *d-p* hybridization are clearly revealed by the spatial charge density. Also, the charge density between W atoms calculated by the MVS approximation is less than that with HSE06, which matches the lower W charge occupation but larger band gap in the MVS approximation. Regarding the second mechanism revealed by the second kind of meta-GGA -TASK and mTASK-, the increased W occupation mainly comes from the enhanced *d-d* interaction between W atoms in the band with VBM. The charge density around Te atoms has no obvious change compared to the PBE.

Overall, the sensitive small band gap in 1T'-WTe$_2$ can be improved through two mechanisms. One is decreasing *d-d* interaction and increasing *d-p* orbital hybridization at the same time, which applies to the screened hybrid functional HSE06 and first kind of meta-GGAs. In this mechanism, without SOC interaction, the band gap can increase from -146 meV with PBE to -31 meV with HSE06, and even reaches a positive value with MVS approximation. Another mechanism is the mainly enhanced *d-d* interaction applicable to meta-GGA TASK and mTASK approximations and SOC correction which increased W but decreased Te occupations for HSE06, PBE and all of the meta-GGAs. The latter meta-GGAs can improve the band gap, but still can't make the gap positive even with the aid of SOC, as shown in Figure 1c. One notable finding is, that combining the two mechanisms together (1$^{st}$ mechanism of MVS approximation with enhanced kinetic energy density dependence plus 2$^{nd}$ mechanism of SOC correction), the fundamental band gap calculated reaches 81 meV, which is close to the ARPES-measured band gap of 55±20 meV. Herewith, MVS appears as a particularly successful combination of both mechanisms. The numerically-evaluated derivative of the exchange-only slope $\frac{\partial F_x}{\partial \alpha}\big|_{\alpha=1}$ is -0.2 for TASK[29], obtained from $F_x$ as a function of $\alpha$ at $\alpha = 1$. Interestingly, this value is nearly the same that we obtain from $\frac{\partial F_x}{\partial \alpha}\big|_{\alpha=1}$ of MVS. In other words, MVS shares a significantly increased slope $\frac{\partial F_x}{\partial \alpha}$ with the TASK approximation as an indicator of better band gaps.

**Evaluation of *GW* Method with PBE and Meta-GGA References.** Typically, the many-body perturbation theory *GW* yields a larger band gap than PBE functional, or corrects the negative band gap revealed by PBE to a positive value, which can in some cases in principle further approach the experimental value via the updated eigenvalues to form the Green's function[38,39]. However, previous calculations found that $GW$ with PBE input produces a smaller band gap than the PBE functional for 1T'-WTe$_2$ material. In the above parts, we clarify that meta-GGA MVS approximation can yield a positive band gap in 1T'-WTe$_2$, which can be even close to the ARPES-measured band gap with the consideration of SOC. Here, we analyze the single-shot $GW$ correction (i.e., $G_0W_0$) in 1T'-WTe$_2$ using MVS as a potentially improved reference. As shown in Figure 5, the $G_0W_0$ correction with MVS reference yields a smaller band gap than the $G_0W_0$ correction with PBE reference. However, both $G_0W_0$@PBE or $G_0W_0$@MVS produces a negative band gap, much smaller band gap than the pure PBE- or the positive MVS-based gap. Negative gaps conflict with the reported ARPES results. Hence, we indicate that the single-shot $GW$ method is not suitable for describing 1T'-WTe$_2$, with either PBE or meta-GGA MVS reference. Besides, the phenomena that the $GW$ correction decreases the band gap more than the PBE functional doesn't occur in the sister material 1T'-MoS$_2$[8], but does occur in 1T'-WTe$_2$. For the two materials with 1T' phase transition, the inverted band near the Fermi level is occupied by W/Mo $d$ and Te/S $p$ orbitals atoms. However, in 1T'-WTe$_2$, both the top valence and bottom conduction bands have mixed $d$ and $p$ characters. Only the MVS approximation can avoid the electron/density diffusion into conduction band (please see the zero atomic occupation yielded by MVS in Figure 3). Such band character is much more complex than the band structure of 1T'-MoS$_2$ in which the band inversion turns the conduction band contributed by metal $d$ orbital to VBM and the VBM turns to CBM contributed by the chalcogenide $p$ orbital[8]. The admixing $d$ and $p$ orbitals in the band inversion should be related to the challenge of capturing the band gap in 1T'-WTe$_2$.

## ■ CONCLUSIONS

Using DFT, Wannier function simulations, the screened hybrid functional HSE06, and first-principles many body perturbation theory $GW$, we have investigated the electronic structure of QSH insulator monolayer 1T'-WTe$_2$, along with the GGA and advanced meta-GGAs. In 1T'-WTe$_2$, the negative band gap described by the PBE functional and the $GW$ method can be corrected to a positive value via the non-empirical meta-GGA MVS approximation, without any means of exact exchange, Hubbard $U$, or SOC correction. The meta-GGA MVS provides an alternative to the high-computation-cost hybrid functionals by reasonably describing 1T'-WTe$_2$. Furthermore, we explain how the sensitive band gap in 1T'-WTe$_2$ can be improved through two mechanisms. One is decreasing the *d-d* interaction and enhancing the *d-p* orbital hybridization at the same time, which applies to the hybrid HSE06 and meta-GGA SCAN, r$^2$SCAN and MVS approximations. Another mechanism is the mainly enhanced *d-d* interaction applicable to SOC correction and meta-GGAs of mTASK and TASK. We also reveal that the exchange component of the MVS approximation displays an enhanced dependence on the kinetic energy density. Combining the two mechanisms together, the fundamental band gap calculated by MVS approximation plus SOC correction reaches 81 meV, which is close to the ARPES-measured band gap of 55±20 meV and the HSE06-calculated gap of 69 meV. Besides, although the $G_0W_0$ correction with MVS reference yields a smaller band gap than $G_0W_0$ with PBE reference, either $G_0W_0$@PBE or $G_0W_0$@MVS produces a much smaller band gap than the pure PBE or MVS case, implying that the GW method is not suitable for describing 1T'-WTe$_2$. Overall, the complexity of the electronic structure of 1T'-WTe$_2$ demonstrates the importance of non-empirical meta-GGAs, especially the MVS approximation, for exploring 1T'-WTe$_2$ and related QSH family materials.

## ■ ACKNOWLEDGEMENTS

This work was supported by the donors of ACS Petroleum Research Fund under New Directions Grant 65973-ND10. AR served as Principal Investigator on ACS PRF 65973-ND10 that provided support for HT. LY and AR acknowledge support from Tulane University's startup fund, which provided support for LY. This research used resources of the National Energy Research Scientific Computing Center (NERSC), a Department of Energy Office of Science User Facility under Contract No. DE-AC02-05CH11231. Helpful discussions with John P. Perdew are acknowledged.

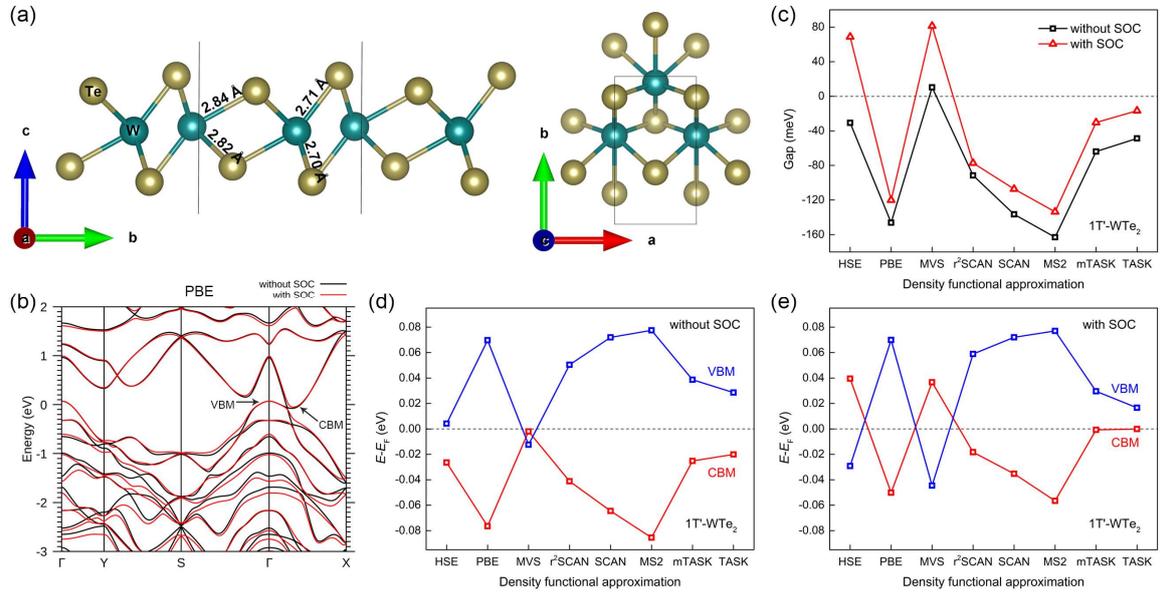

**Figure 1.** (a) The relaxed structure of monolayer 1T'-WTe$_2$ supercell. The cyan and brown balls represent W and Te atoms, respectively. (b) The band structures of monolayer 1T'-WTe$_2$ calculated with PBE and SOC. (c) The calculated band gaps of monolayer 1T'-WTe$_2$ using hybrid functional HSE06, PBE, and different kinds of meta-GGAs, without and with SOC. (d, e) The energy position of valence band maximum (VBM) and conduction band minimum (CBM) of monolayer 1T'-WTe$_2$ using HSE06, PBE and different kinds of meta-GGAs, (d) without and (e) with SOC. The Fermi level in (b), (d), and (e) is located at 0 eV.

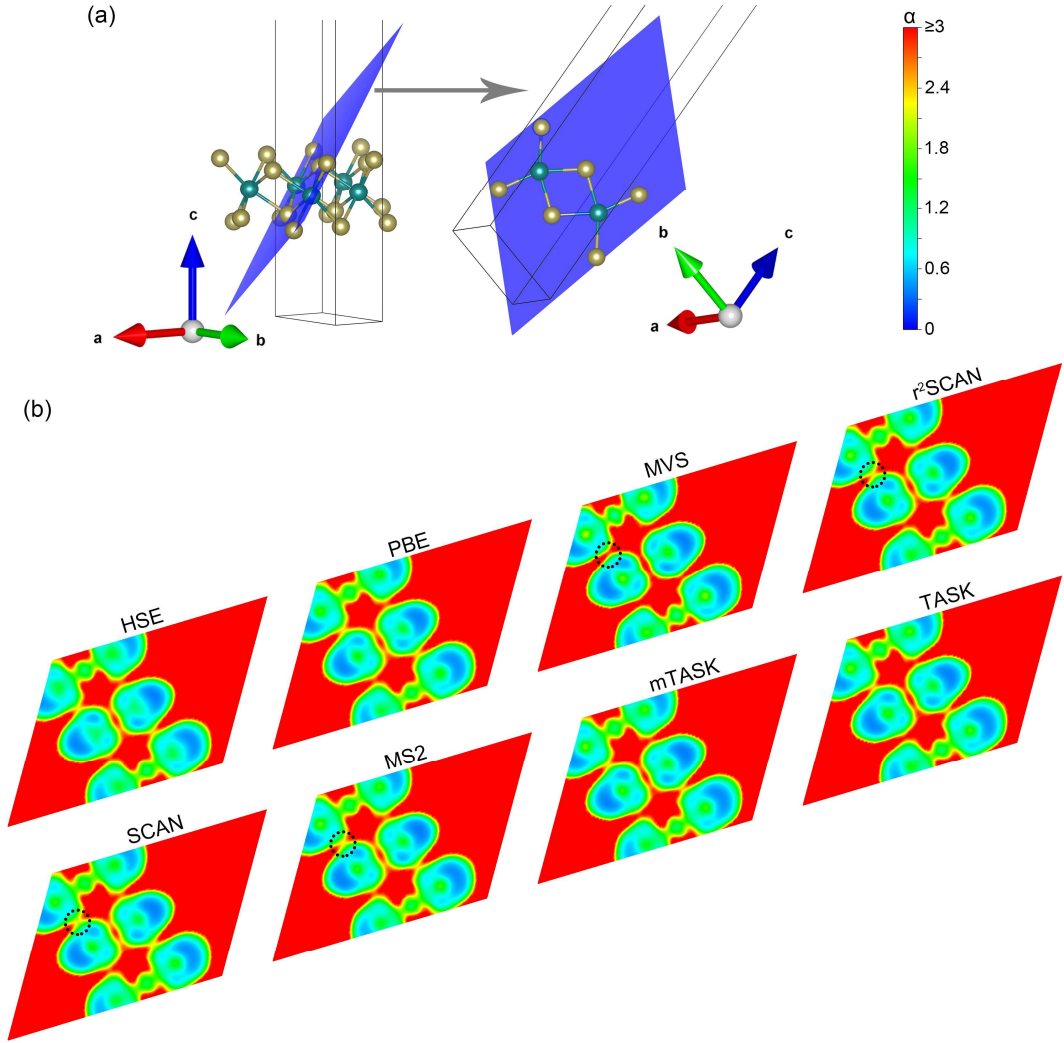

**Figure 2.** (a) The spatial display of the (-1 1 -8) plane in monolayer 1T'-WTe$_2$. (b) The distribution of $\alpha$ in equation (2) in the specific plane displayed in (a), based on hybrid functional HSE06, PBE, and different kinds of meta-GGAs. SOC is not considered. The $\alpha$ value is depicted as the color scale. $\alpha$ approaches $\infty$ in the vacuum limit but is closer to 1 on the Te atoms.

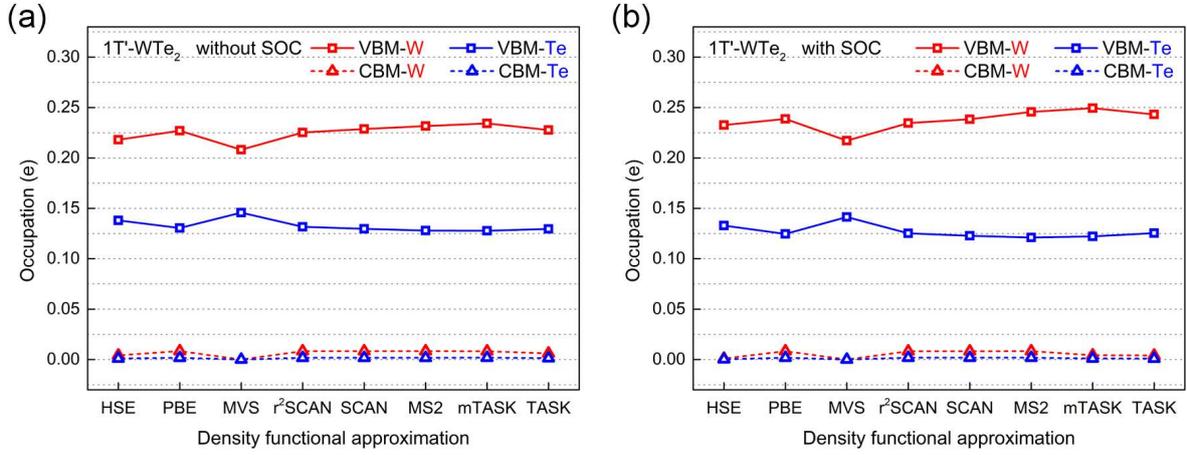

**Figure 3.** The atomic occupations of monolayer 1T'-WTe$_2$ in the band with valence band maximum (VBM) or conduction band minimum (CBM), with hybrid functional HSE06, PBE, and different kinds of meta-GGAs. See the discussion around equation (4). Cases without and with SOC are depicted in (a) and (b) respectively.

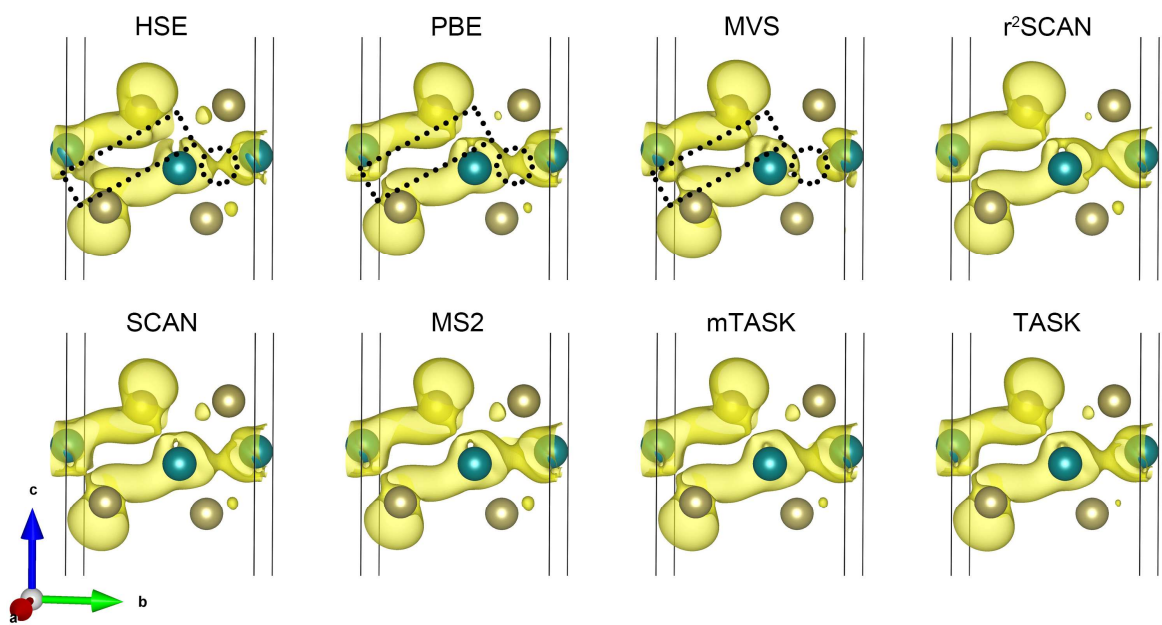

**Figure 4.** The spatial distribution of partial charge density for monolayer 1T'-WTe$_2$ from the band with the valence band maximum, with hybrid functional HSE06, PBE, and different kinds of meta-GGAs. SOC is not included. The displayed isosurface denotes 0.0035 e/Bohr$^3$.

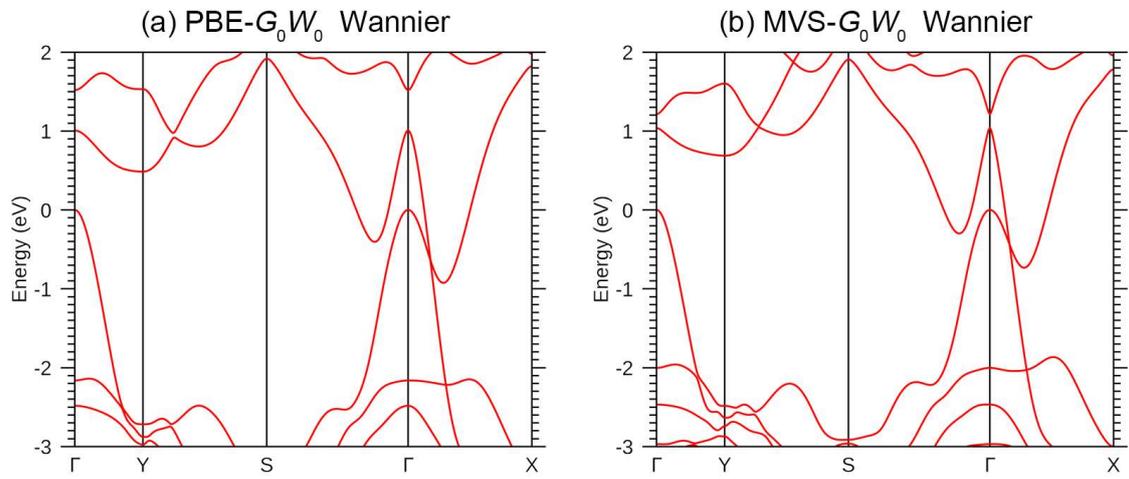

**Figure 5.** The band structures of monolayer 1T'-WTe$_2$ calculated with G$_0$W$_0$ method with (a) PBE and (b) MVS approximations, respectively. The top of the valence band is located at 0 eV. SOC is not included.

# Supplementary Materials

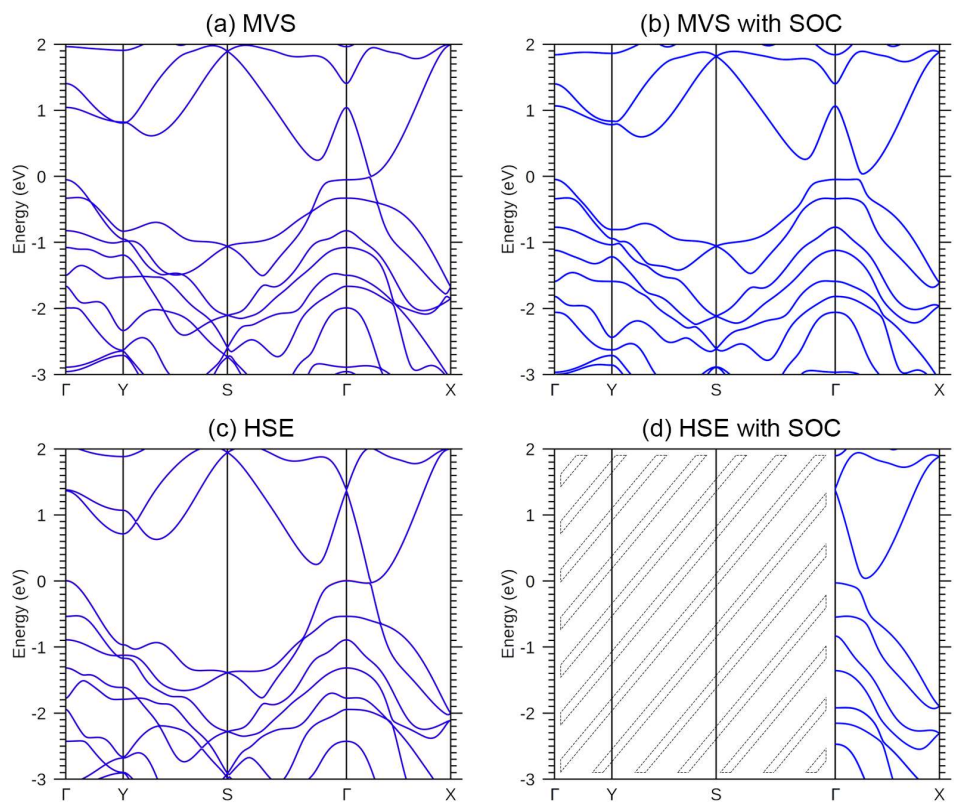

**Figure S1.** The (a, b) MVS- and (c, d) HSE06-based band structures without and with SOC.

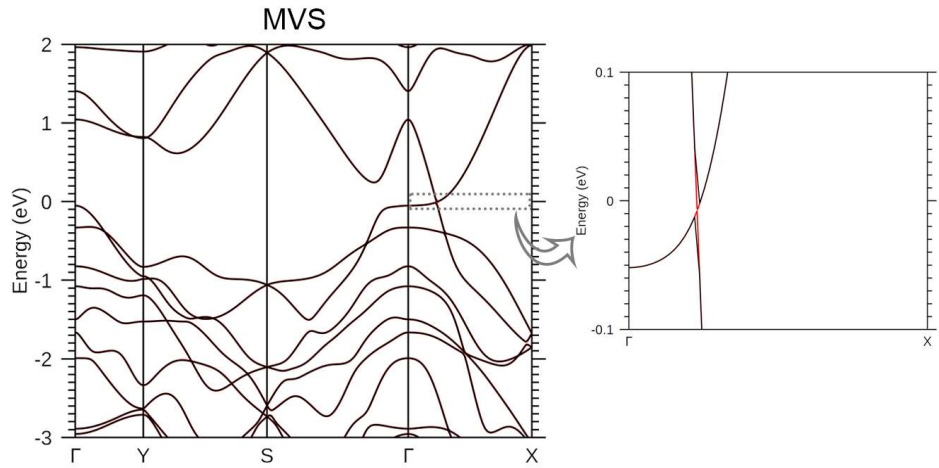

**Figure S2.** The calculated band structures of monolayer 1T'-WTe$_2$ with MVS approximation without SOC. The black color denotes the bands calculated by density functional theory, and the red color curves represent the bands from diagonalizing the Wannier function based Hamiltonian. The right figure corresponds to the dash marked region. The Fermi level is located at 0 eV.

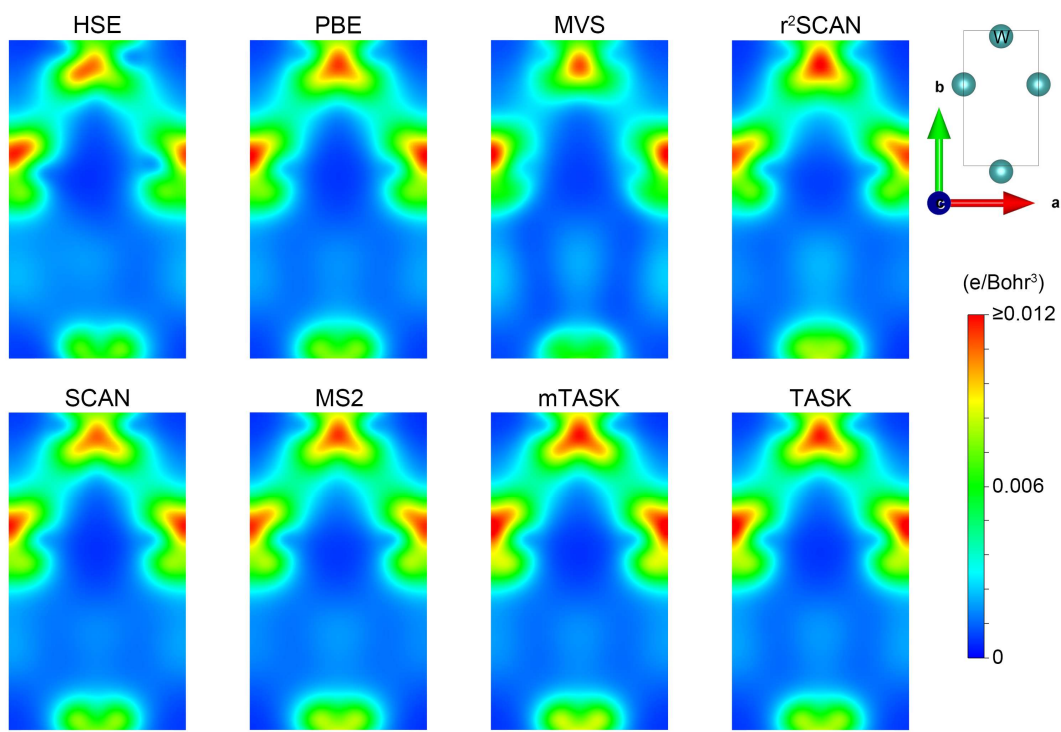

**Figure S3.** The distribution of partial charge density for the (001) planar W layer in monolayer 1T'-WTe$_2$ from the band with the valence band maximum, with hybrid functional HSE06, PBE, and different kinds of meta-GGAs. SOC is not included. The corresponding atomic geometry is shown in the upper right corner. The partial charge density value is coded via the color bar. Note the smaller density on and between the W atoms for HSE06 and MVS, in comparison with PBE.